\documentstyle[aps,multicol,prl,psfig]{revtex}

\draft

\begin{document}

\newcommand{\be}{\begin{eqnarray}}
\newcommand{\ee}{\end{eqnarray}}

\title{Subdiffusive fluctuations of ``pulled'' fronts with multiplicative noise
}

\author{Andrea Rocco$^{1}$, Ute Ebert$^{2}$, and Wim van Saarloos$^{3}$}

\address{$^{1}$Departament ECM, Facultat de F\'{\i}sica, Universitat de 
Barcelona, Av. Diagonal 647, E-08028 Barcelona, Spain}
\address{$^{2}$Centrum voor Wiskunde en Informatica, Postbus 94079, 1090 GB 
Amsterdam, The Netherlands}
\address{$^{3}$Instituut--Lorentz, Universiteit Leiden, Postbus 9506, 2300 RA 
Leiden, The Netherlands}

\date{May 2, 2000}

\maketitle

\begin{abstract}
We study the propagation of a ``pulled'' front with multiplicative noise
that is created by a local perturbation of an unstable state.
Unlike a front propagating into a metastable state, where a
separation of time scales for sufficiently large $t$ creates a diffusive 
wandering of the front position about its mean,
we predict that for so-called pulled fronts, the fluctuations are
subdiffusive with root mean square wandering $\Delta(t) \sim
t^{1/4}$, {\em not} $t^{1/2}$. 
The subdiffusive behavior is confirmed by numerical simulations: For 
$t\le 600$, these yield an effective exponent slightly larger than $1/4$.
\end{abstract}


\begin{multicols}{2}
Since the late 1930s, when the concept of front propagation emerged in the 
field of population dynamics \cite{fisher,kolmog}, interest in this type of 
problems has been growing steadily in chemistry \cite{fife}, physics 
\cite{dee} and mathematics \cite{aw}. In physics, the importance of the 
problem has become more and more clear since it plays a role in a large 
variety of situations, ranging from reaction-diffusion systems to pattern 
forming systems in general \cite{cross}.

Front propagation into unstable states
is an interesting dynamical problem by itself. For a
front evolving from a local perturbation there are but two possible 
propagation mechanisms that are determined by the nonlinearities in
the equation of motion: Either the nonlinearities determine the
velocity of the front that then
is called ``pushed''; or the nonlinearities simply cause saturation 
and the velocity is determined by 
a linearisation about the unstable state. Fronts of this type are called
``pulled'' because they are ``pulled along'' by the spreading and 
growth of small perturbations about the unstable state \cite{ute}. 
Hence pulled front
propagation can occur only if the penetrated state is linearly unstable.
The pushed and pulled regimes are also known as nonlinear and linear 
marginal stability \cite{wimnl}. 
For the discussion below, it is important to realize that pushed fronts relax 
exponentially in time to their long time asymptotes, but that pulled fronts 
relax algebraically without characteristic time scale \cite{ute}.
Hence an adiabatic decoupling of some outer dynamics from 
the internal relaxation of a pulled front is not possible \cite{mba},
and stochastic pulled fronts may show anomalous scaling \cite{KPZ}.

Generally, noise can affect the phenomenological description of a 
reaction-diffusion system in various ways. A first 
possibility is intrinsic noise modelled typically by additive 
thermal noise in a Langevin type equation. 
A second possibility, on which the present paper is focused, 
is at the {\em external} level, e.g. due to fluctuations 
of some control parameter. An example are the
fluctuations of the luminosity intensity in the
photosensitive Belousov-Zhabotinsky 
reaction \cite{sendnad}. Such fluctuations enter
the dynamical equation as multiplicative noise.

The multiplicative noise of the control parameter usually results 
in a modification of the mean propagation velocity of 
the front and in a stochastic wandering of the front position 
around its mean propagation. This means that the noisy front can be 
thought of as a coherent structure whose motion can be decomposed 
into drift plus Brownian motion, very much like a particle 
sedimenting in a fluid. The drift component corresponds to an average 
front, with the average taken over the ensemble of all the realizations 
of the noise. It propagates according to a deterministic equation 
of motion, whose dynamical parameters are in the simplest case just 
renormalised by the noise. 
Theoretically, the important question then arises whether the effects of the 
fluctuations of the front can be understood in terms of a diffusive or 
subdiffusive wandering of some suitably defined front position.

The renormalisation of the front velocity has been studied in the pushed 
and pulled regime \cite{armeroprl}, while the wandering process is 
understood only in the pushed case \cite{armero}, where it has been 
shown to be diffusive: the root mean square position 
of the front $\Delta$ grows with time as $\sqrt{2D_f t}$. Actually, 
the expression for the effective front diffusion coefficient $D_f$
derived by Armero {\em et al.}\ \cite{armero}
was found to break down for pulled fronts, and it was suggested that
the wandering of pulled fronts is subdiffusive.

In this paper we take up the issue of the stochastic wandering 
of pulled fronts about their mean position, and predict 
that in the presence of multiplicative noise pulled fronts behave 
subdiffusively, with $\Delta \sim t^{1/4}$. This prediction is
based on two different arguments. First of all, we heuristically
insert the leading edge asymptotics of the relaxing
pulled front into the expression for the diffusion 
coefficient $D_f$ of pushed fronts, and immediately find
$\Delta \sim t^{1/4}$. Our second argument for the subdiffusive $\Delta 
\sim t^{1/4}$ behaviour comes from mapping the dynamically important region 
onto the KPZ equation. We finally also present data of extensive numerical 
simulations that support our analytical prediction that the wandering is 
subdiffusive with exponent close to 1/4.

The qualitative difference between pushed and pulled fronts 
results from the fact that the 
dynamically important region for {\em pushed} fronts is the interior front 
region, whose extent is finite, while that of {\em pulled} fronts is the 
leading edge ahead of the front \cite{ute}. Starting from a local initial
perturbation, the 
leading edge region grows without bound and as we shall see, this causes the 
subdiffusive behaviour. The power law relaxation of deterministic pulled 
fronts is another manifestation of the leading edge dominated dynamics
of pulled fronts \cite{ute}.

For concreteness, we derive our results by including noise in the one 
dimensional prototype front equation
\be
\frac{\partial \phi}{\partial t} = D \frac{\partial^2 \phi}{\partial x^2} + 
f(\phi)~~~,~~~ f(\phi) = \phi(1-\phi)(a+\phi). \label{def1} 
\ee
Here $a$ is a parameter which plays the role of the control
parameter. Equation 
(\ref{def1}) has a stable state $\phi=1$ and a stationary state $\phi=0$ whose 
relative stability can be tuned by changing the value of the parameter $a$. 
The case $-\frac{1}{2} < a < \frac{1}{2}$ leads to pushed dynamics, while 
$\frac{1}{2} < a < 1$ produces pulled fronts \cite{ute}. For the case
$a=1$, which we will study, the so-called
Fisher-Kolmogoroff-Petrovsky-Piscounoff (F-KPP) equation \cite{fisher,kolmog} 
is recovered.

Let us assume now that the parameter $a$ is replaced by a new fluctuating 
parameter $a(x,t)$ with average $\bar{a}$, $a \rightarrow a(x,t)=\bar{a} + 
\mu(x,t)$, where $\mu(x,t)$ is a Gaussian noise with the moments:
\be
\langle \mu(x,t) \rangle_\mu &=& 0, \\
\langle \mu(x,t) \;\mu(x',t') \rangle_\mu &=& 2 \varepsilon 
\;C(\lambda_{\mu}\;|x-x'|)\;\delta(t-t'), \label{correl}
\ee 
with $\int dx \;C(\lambda_{\mu},|x|) = 1$. We interpret the 
stochastic $p.d.e.$\ defined by (\ref{def1}) -- (\ref{correl}) in the 
Stratonovich sense \cite{Gardiner}. Notice that if $1/\lambda_{\mu}$ 
is much smaller than any other length 
scale in the system, the noise defined by the correlator (\ref{correl}) 
is effectively white in both time and space.

Since according to (\ref{def1}) $\phi$ converges to 1 and is noiseless behind 
the front, we can suitably define the position ${\rm x}_f(t)$ of a noisy front 
propagating to the right into the unstable state $\phi=0$ by
\be
{\rm x}_f(t) = \int_0^{\infty} dx \;\phi(x,t),   
\ee
The displacement $\Delta {\rm x}_f(t)={\rm x}_f(t)-{\rm x}_f(0)$
on average grows with the noise renormalized mean velocity
$\bar{v}_R = \left\langle \dot {\rm x}_f \right\rangle_\mu$.
The fluctuations about the mean displacement $\langle \Delta {\rm 
x}_f(t)\rangle_\mu
=\bar{v}_Rt$ are measured by 
\be
\Delta(t) = \sqrt{\left\langle \left(\Delta {\rm x}_f(t) - \langle
\Delta {\rm x}_f(t)\rangle_{\mu}\right)^2 \right \rangle_{\mu}}.
\ee 
If we relate $\Delta(t)$ to a diffusion coefficient $D_f$ by writing
\be
\Delta^2(t) = \int_0^t dt^{\prime} \;2 D_f(t^{\prime}). \label{diff}
\ee
then for pushed fronts the following expression for the diffusion coefficient 
$D_f$ can be derived \cite{armero,note}:
\be
D_f = \varepsilon \;\frac{\int_{-\infty}^{\infty} d \xi \;
e^{2 \bar{v}_R \xi}\;(d\bar{\phi}/d\xi)^2 \;g^2(\bar{\phi})}
{\left[\int_{-\infty}^{\infty} d \xi \;e^{\bar{v}_R \xi}
\;(d\bar{\phi}/d\xi)^2\right]^2} \label{formula}
\ee
In this formula, $\bar{\phi}$ is the deterministic field associated with the 
front moving with the renormalised pushed velocity $\bar{v}_R$, 
$g(\bar{\phi})=\left.\frac{\partial f}{\partial a} \right|_{\bar{a}}$ is the 
derivative of the reaction term with respect to the control parameter, 
and $\xi={ x}-\bar{v}_R t$ is the comoving coordinate. 

For pushed fronts, 
$D_f$ given by (\ref{formula}) is finite and
time-independent, and hence this gives the diffusive behavior
$\Delta^2(t) = 2 D_f t$. This means that on sufficiently long time scales the 
random displacement is approximately 
Markovian, i.e., the sum of uncorrelated and equally distributed random 
displacements on shorter time scales.

As an example of a pulled front with multiplicative noise, we now study the 
case $\bar{a}=1$:
\be
\frac{\partial \phi}{\partial t} = D \frac{\partial^2 \phi}{\partial x^2} 
+ \phi + \mu \phi - \mu \phi^2 - \phi^3. \label{fishernoise}
\ee
The noise renormalized mean velocity $\bar{v}_R^*$ of the pulled front can be 
calculated explicitly \cite{armeroprl}:
\be
\bar{v}_R^* = \left\langle \dot {\rm x}(t) \right\rangle_\mu
= 2 \sqrt{D(1 + \varepsilon C(0))}. \label{renpulvel}
\ee
However, it is immediately clear that the fluctuation formula 
(\ref{formula}) cannot naively be extended to the pulled regime.

First of all, for a pulled front the expression (\ref{formula}) 
simply diverges. The divergence of solvability-type expressions actually 
holds more generally for perturbative expansions about a pulled front 
\cite{mba}. For a pulled front, 
the dynamically important region is the leading edge defined as the 
region where linearisation about the unstable state is a valid 
approximation; the fact that solvability-type integrals like (\ref{formula}) 
diverge there reflects that the dynamically important region becomes 
semi-infinite.

Second, a pulled front has no characteristic relaxation time \cite{ute},
so there is no reason for the Markovian approximation underlying
diffusive wandering. Rather the leading edge relaxes asymptotically as
\cite{ute}  
\be
\label{steepic}
\phi &\approx& \alpha \xi_R\; e^{-\lambda_R^* \xi_R} 
\;e^{-\xi_R^2/4Dt}/t^{3/2}~~,~~\lambda_R^*=\bar v_R^*/2~,\\
&&~~~\mbox{ for }~ 
\xi_R=x-\bar{v}_R^*t\gg1 ~\mbox{ and }~ t\gg1.
\nonumber
\ee  
The presence of the $\alpha \xi_R/t^{3/2}$ term in front of
the exponentials is actually the fingerprint of the full equation being 
nonlinear. The expression (\ref{steepic}) defines a time dependent Gaussian 
cutoff $\xi_c \sim \sqrt{4Dt}$, which regularizes the integrals in 
(\ref{formula}). In fact, the evaluation of (\ref{formula}) 
with (\ref{steepic}) yields
\be
D_f(t) \approx \frac{3 \varepsilon}{(\bar{v}_R^*)^2 \sqrt{\pi D}} 
\;\frac{1}{\sqrt{t}} \;\;\;\;\; (t \gg 1). \label{formreg}
\ee
Notice that for large times $D_f(t)$ vanishes, marking 
the nondiffusive wandering of pulled fronts. Insertion into 
(\ref{diff}) yields
\be \label{res}
\Delta(t) = \sqrt{2 \int_0^t dt^{\prime} \; D_f(t^{\prime})} \approx 
\left(\frac{12 \varepsilon}{(\bar{v}_R^*)^2 \sqrt{\pi D}}\right)^{1/2} 
t^{1/4}, 
\ee
so the fluctuations are subdiffusive with exponent $1/4$ rather than
$1/2$. 

Although the above argument does capture the essential features 
of fluctuating pulled fronts, it is not entirely systematic, as it 
is based on the extrapolation of the solvability condition 
(\ref{formula}) to the pulled regime.

In order to substantiate the scaling $\Delta(t)\sim t^{1/4}$ 
for a relaxing pulled front with a time-dependent analysis, 
let's go back to Eq.\ (\ref{fishernoise}). The leading edge region 
can be studied by means of the leading edge transformation,
\be
&&\phi(x,t)=\psi(\xi,t)\;e^{-\lambda^* \xi}, \label{letransf}\\
&& \qquad \xi = x-v^*t ~~,~~ v^*=2 ~~,~~ \lambda^*=1. \nonumber
\ee
Eq. (\ref{fishernoise}) can then be written as
\be
\frac{\partial \psi}{\partial t} &=& D\frac{\partial^2 \psi}{\partial \xi^2} - 
\psi  \\
&+& e^{\xi}\left[(1+\mu)\psi e^{-\xi} -\mu \psi^2 
e^{-2\xi} - \psi^3 e^{-3 \xi}\right]. \label{lernl}
\nonumber
\ee 
For $\xi \gg 1$, the nonlinearities can be neglected
\be
\frac{\partial \psi}{\partial t} = D \frac{\partial^2 \psi}{\partial \xi^2} + 
\mu \psi, \;\;\;\;{\rm for} \;\;\;\; \xi \gg 1.\label{linear}
\ee
Notice that the noise in this ``directed polymer'' equation still is 
multiplicative. The Cole-Hopf transformation
\be
\psi(\xi,t)= e^{h(\xi,t)}, \label{kpztransf}
\ee
converts (\ref{linear}) into an equation with additive noise:
\be
\frac{\partial h}{\partial t} = D \frac{\partial^2 h}{\partial \xi^2} + 
D \left(\frac{\partial h}{\partial \xi}\right)^2 + \mu, \;\;\;\;{\rm for} 
\;\;\;\; \xi \gg 1.\label{kpz}
\ee
Eq.\ (\ref{kpz}) is the celebrated 1-dimensional Kardar Parisi Zhang (KPZ) 
interface equation \cite{parisi}.

The essential difference between our problem and previous studies
of the KPZ equation are the initial and boundary conditions. 
After some temporal evolution, the nonlinearities in the original 
$\phi$ equation will lead to the fluctuationless saturation 
of $\phi$ at the value of unity for $\xi \ll -1$, 
which corresponds to the fluctationless slope $h \approx \lambda^* \xi$ 
behind the front: It is as if  the KPZ equation has to be solved in the
positive half-space with (roughly) a fixed boundary. On the other hand,
by translating   (\ref{steepic}) back into $h$, we see that for large
$\xi$ and $t$, the average interface shape $h_{av}$ should be given by   
\be
h_{av} \approx \ln(\alpha \xi_R/t^{3/2}) +\lambda^*\xi -\lambda^*_R
\xi_R  - \xi_R ^2/4Dt~~.
\ee
Thus, apart from the logarithmic term the average interface is
essentially tilted but flat up to  the time-dependent cross-over 
$\xi_c \approx \sqrt{4Dt}$ \cite{note2}, and beyond $\xi_c$ it
has the shape of a downward curved parabola with time dependent curvature.
Together with the fact that the nonlinear term
in (\ref{kpz}) gives an average nonzero growth velocity, this makes the
problem into a nonstandard fluctuating interface problem. Our central 
approximation is now to consider the relaxing front in the essentially 
straight but fluctuating section between 0 and $\sqrt{4Dt}$ as a KPZ 
interface with time dependent length $L={\cal O}(\xi_c)$. As the scaling 
exponents of the KPZ equation are robust with respect to a geometric change 
of the fluctuating surface \cite{spohn}, we use the KPZ scaling functions
for the root mean square width $W$ of the interface $h$,
\be
W(L,t)= t^{\beta}Y\left(\frac{t}{L^z}\right), \;\;\beta=1/3,\;\;z=3/2,
\ee
where $W=\sqrt{\langle
\overline{h(x,t)-\overline{h}(x,t))^2}\rangle_\mu}$, with the bar
denoting a spatial average. The scaling function $Y(s)$ will 
depend on the shape of the roughening surface, but always has the limits 
$Y(s) \rightarrow s^{-\beta}$ for $s \rightarrow \infty$, $Y(0)
\approx {\rm const}$.

Inserting our approximation $L \sim \sqrt{t}$, we get:
\be
W(L,t) \sim L^{z\beta} \sim (\sqrt{t})^{z\beta}= t^{1/4}. 
\ee
The final step of our argument is to convert this result in a prediction for 
the fluctuations of the front position. If we measure the position of the 
front by tracking a certain height $c$, $\phi(x_c,t) = {\rm const} = c$, 
and use the relations (\ref{letransf}) and (\ref{kpztransf}), we find:
\be
\phi(x_c,t)=e^{-\lambda_R^*(x_c - \bar v_R^* t) + h}={\rm const} = c. 
\ee
This implies that fluctuations in $h$ are just identical with fluctuations in 
$x_c$. 
Therefore we get
\be
\Delta(t) \sim t^{1/4} 
\ee
which reproduces the scaling of our previous result (\ref{res}).

We have also performed numerical simulations of the noisy front equation 
(\ref{def1}) with $a=-0.3$ (pushed) and $a=1$ (pulled, F-KPP 
Equation (\ref{fishernoise})) following the lines of \cite{armero}. 
The initial condition was taken as a step function $\phi(x,0)=\theta(x_0-x)$.
The numerical integration has been performed 
using a standard explicit Euler algorithm, in both cases the value of the 
noise was set to $\varepsilon=0.5$, and the zero value of the spatial noise 
correlator $C(0)$ was chosen as the inverse spatial integration
mesh, $C(0)=1/\Delta x$ \cite{armero}. The result is shown in Fig.~1, 
where the function $\Delta(t)$ is plotted in both the pushed and 
the pulled case.
 
The specific features of the pulled regime make the problem quite delicate 
from the numerical point of view. In order to minimize finite size
effects, which are particularly worrisome in this regime \cite{ute}, 
we have worked with a large system size ($L=3000$) and gridsize 
$\Delta x=1$ (the change in $v^*$ and $D$ due to the finite gridsize effect 
was taken into account following the prescription of \cite{ute}). 
This made sure that even at time $t=600$, the leading edge of the front 
never reached the boundary of the system. 

We have also checked our program and 
system size extensively both for deterministic and noisy fronts, taking into 
account grid and time step effects according to \cite{ute}.

\begin{figure}[h]
\centerline{{\psfig{figure=fig_resub.eps,angle=-90,height=6cm}}}
\end{figure}


\noindent
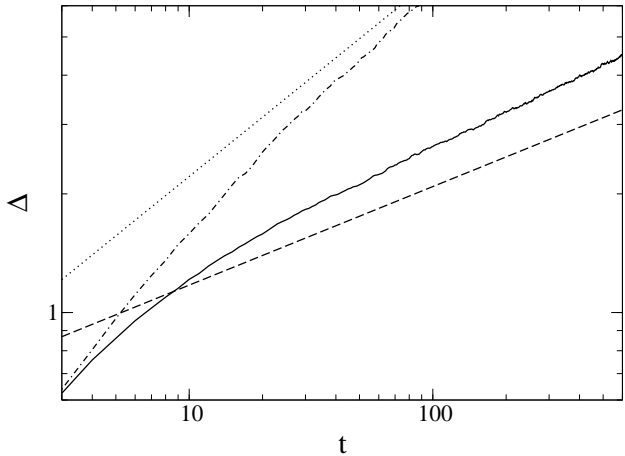
FIG. 1. {\small{Diffusive and subdiffusive spreading of the front position. 
The dot-dashed curve correponds to the pushed case ($a=-0.3$) and the 
solid one corresponds to the pulled case ($a=1$). The dashed straight
line is the prediction (\ref{res}), while the dotted line indicates a slope 
$1/2$.}}

\vspace{.2cm}

Our final result, based on averaging over 10000 front realizations, 
is shown in Fig.~1; it clearly  confirms the subdiffusive 
behaviour predicted by our analytical arguments. 
Quantitatively, 
when we associate a single effective exponent with the late time slope in 
the log-log plot of Fig.~1, we get an effective exponent of about 0.29
rather then $1/4$. Over the time interval we have studied, the actual
value of $\Delta(t)$ is somewhat larger than an asymptotic prediction
(\ref{res}), which is indicated with a dashed line. This may be due to
the fact that (\ref{res}) only gives the behavior for such long times
that the time integral is dominated by its large $t$ behavior. The
fact that $\Delta$ is only of the order of $4$ at our latest times
suggests that this asymptotic regime is only reached at very late
times. Indeed, assuming that finite size effects are negligible, we attribute
the fact that the effective exponent is slightly larger than $1/4$ 
to the presence of slow crossovers, which surely are 
present in the system. Some of these can be estimated, while others are more
difficult to trace. (i) We already noticed previously that we are actually 
dealing with a slightly curved KPZ interface, for which the crossover scaling 
functions are not known, and that the way in which the cutoff 
$\xi_c = {\cal O}(\sqrt{t})$ enters the KPZ analysis requires further 
study. (ii) The corrections to our asymptotic estimates for the integrals in 
(\ref{formula}) are all of order $1/\sqrt{t}$, with possible logarithmic 
corrections \cite{ute}. This indicates that the corrections to the scaling
$\Delta \sim t^{1/4}$ are of order $t^{-1/4}$, possibly with logarithmic 
corrections. (iii) If initially $\phi$ falls off as $\exp(-\lambda_R^* x)$, 
then the associated KPZ interface remains straight towards $\xi=\infty$. For
this case the KPZ scaling predicts $\Delta \sim t^{1/3}$. Presumably
a crossover between exponent $1/3$ and $1/4$ could be present when 
starting with an initial condition slightly faster decaying than 
$\exp(-\lambda_R^* x)$. The identification of such a crossover and the 
modification of the global exponent due to these special initial conditions 
is an issue that will be addressed elsewhere.

We finally stress, that our results apply to a much larger class
of equations than nonlinear diffusion equations (\ref{def1}).
The methods of generalization are analogous to those of \cite{ute,mba};
a closely related result is the general argument put forward in 
\cite{KPZ} that noisy pulled fronts in more than one dimension should 
not obey KPZ scaling.

We thank J. Casademunt and L. Sch\"afer for useful discussions. 
A.R. thanks the Instituut-Lorentz for kind hospitality.
He was supported by the European Commission project ERBFMRX-CT96-0085 
and U.E. by the Dutch Science Foundation NWO.

\end{multicols}


\begin{thebibliography}{99}
\bibitem{note} In \cite {armero}, Eq. (\ref{formula}) was arrived at by a 
solvability type analysis assuming a separation of time scales. We
will improve this procedure elsewhere, but the more careful analysis 
does indeed yield the same result (\ref{formula}).
\bibitem{note2} How to obtain this cutoff selfconsistently within 
the KPZ formulation is not completely clear; it obviously is associated with 
the fact that the terms $\partial^2 h/\partial \xi^2$ and 
$(\partial h/\partial \xi)^2$ are both of the order $1/t$ at distances $\xi$
of order $\sqrt{t}$.
\end{thebibliography}
\end{document}